\documentstyle[aps,prl,floats,epsfig]{revtex}

\newcommand{\eins}{\mbox{$1 \hspace{-1.0mm}  {\bf l}$}}

\newcommand{\be}{\begin{equation}}
\newcommand{\ee}{\end{equation}}
\newcommand{\bea}{\begin{eqnarray}}
\newcommand{\eea}{\end{eqnarray}}

\newcommand{\ket}[1]{ | \, #1  \rangle}
\newcommand{\bra}[1]{ \langle #1 \,  |}

\newcommand{\proj}[1]{\ket{#1}\bra{#1}}

\newcommand{\calw}{\mbox{$\cal W$}}

\newcommand{\rhota}{\mbox{$\rho^{T_A}$}}

\newcommand{\fc}[1]{f^{\dagger}_{ #1 }}
\newcommand{\0}{\Omega}
\newcommand{\calc}{\mbox{$\cal C$}}
      
\newcommand{\C}{\mbox{$\cal C$}}

\begin{document} \draft

\title{Entanglement properties of composite quantum systems}

\author{
K. Eckert, O. G\"uhne, F. Hulpke, P. Hyllus, J. Korbicz,
J. Mompart,
D. Bru\ss , M. Lewenstein, and A. Sanpera
\\
{\em Institut f\"ur Theoretische Physik,
Universit\"at Hannover,
Appelstr. 2, D-30167 Hannover, Germany }\\
}
\date{Received \today}
\maketitle

\begin{abstract}
We present here an overview of our work concerning
entanglement properties of composite quantum systems. 
The characterization of entanglement, i.e. the 
possibility to assert if a given quantum state is entangled
with others and how much entangled it is, remains one of
the most fundamental open questions in quantum information theory. 
We discuss our recent results related to the problem of separability
and distillability for distinguishable particles, employing the
tool of witness operators. Finally, we also state our results concerning
quantum correlations for indistinguishable particles.
\end{abstract}

\section{Introduction}

The processing of quantum information differs in
a fundamental way
from the processing of classical information: rather than allowing
only boolean values ``0'' and ``1'' for a bit, 
a quantum bit or qubit is implemented by the quantum state of a two-level
system, which can  be
in any superposition of $``\ket{0}$'' and $``\ket{1}$'', namely
$\ket{\psi}=\alpha \ket{0}+\beta \ket{1}$. If several
quantum states are involved, rather than dealing only with one string of
bit values, e.g., ``001011'', the state of the composite system of qubits
can be in a superposition of such strings, e.g.,
$\ket{\Psi}=a \ket{001011}+b\ket{110100}+c\ket{010010}+....$. 
In general such a state cannot be written as a tensor product of
states of its subsystems and, therefore, it is called {\em entangled}.

Entanglement is a key feature for most of the protocols used 
in quantum information such as, e.g., quantum teleportation, quantum
cryptography, superdense coding, quantum algorithms
and quantum error correction. Indeed the resources needed to implement
a particular protocol of quantum information are closely linked to the
entanglement properties of the states used in the protocol. Therefore,
it is highly desirable to characterize the entanglement properties of
quantum systems bearing in mind that this is a fundamental 
open problem of quantum theory but also that it is essential for the
implementation of any possible task that relies on quantum bits.

In this manuscript we summarize our efforts in striving at some
understanding of the properties of entanglement for composite quantum 
systems of distinguishable and indistinguishable particles. 
In the former case, one assumes that the involved quantum systems 
can be addressed separately. This happens either because the subsystems 
are located at different places so that their wavefunctions do not spatially
overlap, or because they differ in some degrees of freedom permitting thus to
distinguish them.  
Until quite recently, this has been the most common approach 
considered in the framework of quantum information. 
However, the experimental progress in 
achieving quantum bits and quantum gates by means of solid state
physics (quantum dots) and optical microtraps with neutral atoms, 
demands a new formalism which includes the statistical nature of the
particles involved. In the last part of this manuscript we address
this question.

In the frame of the DFG-Schwerpunkt on ``Quanteninformationsverarbeitung''
(quantum information processing) we have addressed these subjects 
in two different projects. The manuscript aims to give a comprehensive 
summary of our results for a reader familiar with the subject of 
separability and distillability.
This summary, however, is by no means exhaustive. Important 
contributions to the above projects concerning  entanglement measures, 
catalysis of entanglement and quantum game theory  
are presented elsewhere \cite{potsdam}.

The manuscript is organized as follows: Sections \ref{sep}-\ref{wit} deal with
the characterization of entanglement of composite quantum systems of
distinguishable particles. In section
\ref{sep} we first briefly state the problem of separability, i.e., the 
question when the state of a composite quantum system does not 
contain any quantum correlations or entanglement, and then  
report our results concerning this question. 
In section \ref{distil} we address the
problem of distillability, i.e., the question when the state of 
a mixed composite quantum system can be transformed  to
a maximally entangled pure state by using local operations and classical
communication. We report thereafter our progress concerning this subject. 
Section \ref{wit} deals with witness operators. We state our
achievements in constructing, optimizing and implementing witness
operators to detect entanglement. Also, a connection between a
witness detecting a given state and the distillability and activability 
properties of the state is presented there.
Finally, in section \ref{fer} we address 
the study of quantum correlations in composite systems of 
identical particles and apply some of the formalisms
previously developed for distinguishable particles to 
indistinguishable ones.

\section{ Separability of composite quantum systems}
\label{sep}
In this section, before presenting our results, we define 
the problem of separability versus
entanglement for a given quantum system. The reader interested in 
a tutorial description of the subject is addressed to references  
\cite{primer} and \cite{tutorial}.

For simplicity, let us restrict ourselves here 
to the simplest case of composite systems: bipartite systems 
(traditionally denoted as Alice and Bob) of finite, but 
otherwise arbitrary dimensions. Physical states of such systems are, in
general, mixed and are described by density matrices, i.e. hermitian,
positive semi-definite linear operators
of trace one (i.e. $\rho=\rho^{\dagger}$, $\rho\ge0$,${\rm Tr}\rho=1$), acting in the Hilbert space of the composite system 
$\cal H=\cal {H}_A \otimes \cal {H}_B$.
Without loosing generality we will assume that
dim ${\cal {H}}_A=M\ge 2$ and dim ${\cal {H}}_B=N \ge M$. 

Before proceeding further we introduce here some definitions that we
will use throughout the paper. 
Given a density matrix $\rho$,  we denote its kernel by
$K(\rho)=\{\ket\phi: \rho\ket\phi=0\}$, its range by $R(\rho)=\{\ket\phi:\exists \ket{\psi}\}$, 
and its rank by $r(\rho)={\rm dim}\; R(\rho)=NM-{\rm dim}\, K(\rho)$.
Also the notion of {\em partial transposition} will be used
throughout. The operation of partial transposition of a density 
matrix $\rho$ means the 
transposition with respect to only one of the subsystems. If we express $\rho$ in
Alice's and Bob's orthonormal product basis,
\be
\rho=\sum_{i,j=1}^{M}\sum_{k,l=1}^{N}\bra{i,k} \rho\ket{j,l}\ket{i,k}\bra{j,l}
=\sum_{i,j=1}^{M}\sum_{k,l=1}^{N}\bra{i,k} \rho\ket{j,l}{\ket{i}}_A\bra{j}\otimes
{\ket{k}}_B\bra{l},
\ee
then, the partial transposition with respect to Alice's system is given by:
\be
\rho^{T_A}=\sum_{i,j=1}^{M}\sum_{k,l=1}^{N}\bra{i,k} \rho\ket{j,l}\ket{j}_A
\bra{i}\otimes\ket{k}_B\bra{l}.
\ee
Note that $\rhota$ is basis-dependent, but its spectrum is not.  
For the partial transpose $\rhota$ it might hold that $\rhota\geq 0$, but this
does not have to be true!
As $({\rhota})^{T_B}=\rho^{T}$, and as $\rho^{T}\geq 0$ always holds,
positivity of $\rhota$ implies positivity of $\rho^{T_B}$ and vice versa.
A density matrix $\rho$ that fulfills $\rhota\ge0$ is termed
{\em PPT state} for positive partial transpose, otherwise it is called {\em NPPT state}
for non-positive partial transpose.
\subsection{The separability problem}

An essential step towards the understanding of entanglement  
is to first identify separable states, i.e, states that contain classical
correlations only or no correlations at all. 
The mathematical definition of such states (separable states) in terms
on convex combinations of product states was given by Werner in \cite{werner}. 

\noindent {\bf {Def. 1}} A given state $\rho$ is {\em separable}
iff \be \rho=\sum_{i=1}^{k}p_i \rho_i^{A}\otimes \rho_i^B\ ,
\label{separable} \ee where $\sum_ip_i=1$, and $p_i\ge 0$.

Notice that the above definition states that a separable state 
can be prepared by Alice and Bob by means of local operations 
(unitary operations, measurements, etc.) and classical communication
(LOCC). However, the question whether a given state can be decomposed 
as a convex sum of product states like in eq. (\ref{separable}) 
is by no means trivial -- in fact there are no algorithms to check if
such a decomposition for a given state $\rho$ exist.

An entangled state is defined via the negation of the above
definition. 
A given state $\rho$ is {\em entangled} iff it cannot be decomposed
as in Equation (\ref{separable}).
Thus, the separability versus entanglement
problem can be formulated  as:
{\it {Given a composite quantum state described by $\rho$,
can it be decomposed as a convex combination of product states or
not?}}.

A major step in the answer of this problem and in the characterization of
separability  was done by Peres\cite{peres1} and the Horodecki
family\cite{horo} by providing a necessary condition for separability:
the positivity of the partial transposition. Their results can be 
summarized in the following theorem:

\noindent{\bf {Theorem 1}} 
If a density matrix $\rho$ is separable then $\rhota\geq 0$.
If $\rhota\geq 0$ in Hilbert spaces of dimensions $2\times 2$ or
$2\times 3$ then $\rho$ is separable.

Notice that being PPT does not imply separability, except for low dimensional Hilbert spaces!
Let us mention here, that also in \cite{horo}, 
the problem of separability was rigorously reformulated 
in terms of the theory of positive maps. We will discuss about
positive and completely positive maps in the forthcoming sections.
 
\subsection{Results on the separability problem}

\begin{figure}[ht]
\setlength{\unitlength}{1pt}
\begin{center}
\psfig{file=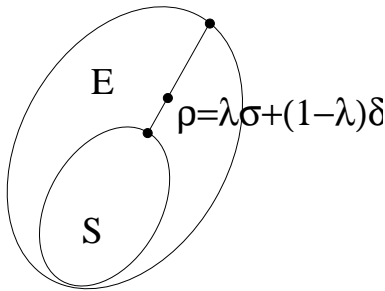,width=5cm}
\end{center}
\caption[]{\small Illustration of the decomposition of $\rho$ into a separable state $\sigma$ and an edge state $\delta$:
$\rho=\lambda\sigma+(1-\lambda)\delta$}
\label{figdecomp}
\end{figure}

An important tool for studying the properties of states with
respect to their separability is the method of subtracting
projectors onto product vectors from the given state. This method
was developed in \cite{lsd} and \cite{anna}: if there exists a
product vector $|e,f\rangle\in R(\rho)$, the projector onto this
vector (multiplied by some coefficient $\lambda>0$) can be
subtracted from $\rho$, such that the remainder is positive
definite.  A similar technique can be used for PPT states $\rho$:
if there exists a product vector $|e,f\rangle\in R(\rho)$,  such
that $|e^*,f\rangle\in R(\rhota)$, the projector onto this vector
(again multiplied by some $\lambda>0$) can be subtracted from
$\rho$, such that the remainder is positive definite and PPT.
This observation allows to construct decompositions of a given
$\rho$ of the form
\be
\rho=\lambda\sigma +(1-\lambda)\delta,\label{dec}
\ee
where $\sigma$ is separable, while $\delta$ is a so-called
 {\it edge state},
i.e. a state from which ``nothing else'' can be subtracted.
In the case of decompositions for general entangled states, $\delta$
has no product vectors in the range. In the case of decompositions
for PPT states  $\delta$ can be taken as PPT edge state, i.e.
a state that does not contain any product vector $|e,f\rangle\in R(\rho)$,  such that
$|e^*,f\rangle\in R(\rhota)$.
The decompositions (\ref{dec}) can be optimized,
 by demanding $\lambda$ to be maximal \cite{lsd}. Such an optimal
 decomposition is illustrated in Figure \ref{figdecomp}.
For separable states an important question (related to the
optimization of the detection of entangled states
\cite{detection}) concerns minimal decompositions, i.e. those
containing minimal number of projectors on product vectors. In
particular in Ref. \cite{anna} it has been shown that in
$2\times2$ systems the minimal decomposition of separable states
contains a number of projectors which is equal to the rank of the
state. Minimal decompositions can also be considered in the form
of pseudo-mixtures, where not all coefficients multiplying the
projectors entering the decomposition are positive, see section
\ref{reswit}.

In the following we list the major results obtained by applying the
decompositions (\ref{dec}) and constructing edge states
in various systems, which has become a basic tool of
the so-called Innsbruck-Hannover programme \cite{reflections}.

 General properties of optimal separable approximations
(decompositions) have been studied in Ref. \cite{sepap} for the
states $\rho$ of bipartite quantum systems of arbitrary dimensions
$M\times N$.
 For two qubit systems (M=N=2) the best separable
       approximation has a form of a mixture of a separable state and
a projector onto a pure entangled state. We have formulated the
necessary condition that the pure state in the best separable
approximation is not maximally entangled. This result allowed
Wellens and Ku\'s \cite{wellens} to obtain an analytic form of the
optimal decomposition in the $2\times2$ case and to relate the
value of $\lambda$ to the
Wootters'
concurrence\cite{wooters}.
 We have
       demonstrated that the weight of the entangled state in
the best separable approximation in arbitrary dimensions provides
a good entanglement measure. We have proven that
 in general, for arbitrary M and N,  the best separable
       approximation corresponds to a mixture of a separable and
an entangled state which are both unique. We have developed
 also a theory of optimal separable approximations for states
with positive partial transpose, and
 discussed procedures of constructing such decompositions.

The decomposition techniques and investigations of edge states
have then be applied to $2\times N$ systems in \cite{onlyweb} and
\cite{lowrank}. We have analyzed the separability properties of
PPT density operators supported on $\C^2\otimes \C^N$. We have
shown
 that if $r(\rho)=N$, then it is separable, and that bound
       entangled states have rank larger than $N$.
We have also solved the separability problem for low rank states:
we have  given a separability criterion for a generic density
operator such that the sum of its rank and the one of its partial
transpose does not exceed $3N$. If it exceeds this number we
       show that one can subtract projectors onto
product vectors until decreasing it to $3N$, while keeping the
positivity of $\rho$ and its partial transpose. This automatically
gives us a sufficient criterion for separability for general
density operators.
       We also prove that all density operators that remain invariant (or, more generally close
       to being invariant)
after partial transposition with respect to the first system are separable.
Finally, in Ref. \cite{onlyweb} we have also presented  a simple elementary proof
of the Peres-Horodecki separability criterion in $2\times 2$--dimensional
systems.

The results for $2\times N$ systems were then generalized to
$M\times N$ systems\cite{low2}, where we have also been able to
solve the separability problem for  low rank states. We have
considered low rank density operators $\varrho$ supported on a
$M\times N$ Hilbert space for arbitrary  $M\leq N$ and with a
positive partial transpose  $\varrho^{T_A}\ge 0$. For rank
$r(\varrho) \leq
       N$ we have proven that having a PPT is necessary
and sufficient for $\varrho$ to be separable; in this case we have
also provided its minimal decomposition in terms of pure product
states. It follows from this result that there are no
       bound entangled states of rank $3$ having a PPT.
We have also presented a necessary and sufficient condition for
the separability of generic density matrices for which the
sum of the ranks of $\varrho$ and $\varrho^{T_A}$ satisfies
       $r(\varrho)+r(\varrho^{T_A}) \le 2MN-M-N+2$.
This separability condition has the form of a constructive check, providing thus also a pure product state decomposition for separable states,
and it works in those cases where
       a system of coupled polynomial equations
has a finite number of solutions, as expected in the generic case.

The same research programme can also be applied to $2\times
2\times N$ systems\cite{22n}. We have investigated separability
and entanglement of mixed states in ${\cal C}^2\otimes{\cal
C}^2\otimes{\cal C}^N$ three-party quantum systems. We have shown
 that all states $\rho$  with positive partial transposes that have rank $r(\rho)\le N$
       are separable. For the three-qubit case (N=2) we have proven
 that all PPT states $\rho$ that have positive partial transposes and rank
$r(\rho)=3$ are separable.   We provided also constructive
separability checks for the states $\rho$ that have the sum
       of the rank of $\rho$ and the ranks of partial transposes
with respect to all subsystems smaller than 15N-1.

 We have studied also the problem of separability and
entanglement properties of completely positive maps acting on operators
acting in the
composite Hilbert space of Alice and Bob\cite{rare,forts}. We have studied
 when a physical
operation can produce entanglement between two systems
that are initially
disentangled. The formalism that we have
developed allows to show that one can perform certain
non-local operations with unit probability by
       performing local measurements on states that are very weakly entangled.
This formalism is a generalization of the Jamio\l kowski
isomorphism, that connect maps with operators, to the case of maps
acting on tensor product spaces. We have associated with every
completely positive map (CPM) acting on states of Alice and Bob,
an operator acting on two copies of  Alice's and Bob's space. The
isomorphism connects separable CPMs with separable states, PPT
CPMs with PPT entangled states, and so one. It provides a powerful
tool to classify CP maps, using results known for states.

Last,  but not least, we have applied our methods and techniques
to study states in infinite-dimensional Hilbert spaces, i.e.
continuous variable states. A particularly important class of such
states, that is very frequently used in experiments with photons,
is formed by the so-called ${\it Gaussian}$ states. Gaussian
states can be defined by the requirement that their associated
Wigner function has a Gaussian form. We have been able to solve
the separability problem of Gaussian states for two parties each
having an arbitrary number of photon (harmonic oscillator) modes
(\cite{sepgauss}, for a review see \cite{forts2}), and for three
parties each having one harmonic mode\cite{gaussian}. For
bipartite systems of arbitrarily many modes  the necessary and
sufficient condition consists in an iterative transformation of 
the correlation matrix of a
given state and provides an operational criterion, since it can be
checked by a simple computation with arbitrary accuracy.
Moreover, it allows us to find a pure
product-state decomposition of any given separable Gaussian state.
Our criterion is independent of the one based on partial transposition,
and obviously, since it detects all entangled states, it is
 strictly stronger than the PPT criterion. We have also
derived a necessary and sufficient condition for the separability
of tripartite three mode Gaussian states, that is easy to check for any
such state. We have given a classification of the separability properties of
those systems and have shown how to determine for any state to which
class it belongs.
We have also shown that there exist genuinely tripartite
bound entangled states (see \ref{distil})
and have pointed out how to construct and prepare such states.

\section{The distillability problem}
\label{distil}

For many applications in quantum information processing
one needs a maximally
entangled state of two parties, i.e. a state
in $M\times N$ dimensions of the form
\be
\ket{\Psi_{max}}=\frac{1}{\sqrt{M}}\sum_{i=1}^{M}\ket{i,i}\ .
\ee
However, even if an experimental source that creates such a state
is available, during storage or transmission along a noisy channel
the state will interact with the environment and evolve into
a mixed state, thus loosing the property of being maximally
entangled.

The idea of distillation and purification, i.e. enhancement of the
entanglement of a given mixed  state by local operations and
classical communication (LOCC) was proposed by Bennett {\it et
al.} \cite{bennet}, Deutsch {\it et al.}~\cite{deutsch} and Gisin
~\cite{gisin}. Again, for Hilbert spaces of composite systems with
dimension lower or equal to 6, any mixed entangled state can
always be distilled to a pure maximally entangled state. Since for
such systems entanglement is equivalent to  non-positivity of the
partial transpose, we conclude that for systems in $2\times 2$ and
$2\times 3$ dimensions all NPPT states are distillable
\cite{lowdis}. It was shown by the Horodecki family \cite{bound}
that the PPT property implies undistillability.
Somehow surprisingly, in higher
dimensions there exist states that are entangled but cannot be
distilled. These states, namely PPT entangled states, are
called bound entangled states, contrary to free entangled
states which can be distilled.
%
%
In general, the distillability  problem can be formulated as:
{\it{ Given a composite quantum state described by $\rho$, is it
distillable or undistillable?}}

The problem of distillability can be rigorously formulated \cite{bound}
so that it reduces to the following theorem:

\noindent{\bf {Theorem 2}}
$\rho$ is distillable iff there exists a state $\ket{\psi}$ from a
$2\times 2$-dimensional subspace,
$
\ket{\psi}=a\ket{e_1}\ket{f_1}+b\ket{e_2}\ket{f_2}\ ,
\label{psi2by2}
$
such that 
\be
\bra{\psi}(\rho^{T_A})^{\otimes K}\ket{\psi}<0
\ee for some $K$.

\subsection{Results on the distillability problem}

As mentioned above, in finite dimensions there exist so-called
{\em bound entangled} states, which are entangled, but their
entanglement cannot be distilled. One possibility of the
construction of such states was given in \cite{chess}. There we
have presented a  family of bound entangled states in $3\times3$
dimensions. Their density matrix
     depends on 7 independent parameters and has  4 different
     non-vanishing eigenvalues. This construction can e.g. be useful
     when testing whether some new entanglement criterion
     detects bound entanglement.

\begin{figure}[ht]
\setlength{\unitlength}{1pt}
\begin{center}
\psfig{file=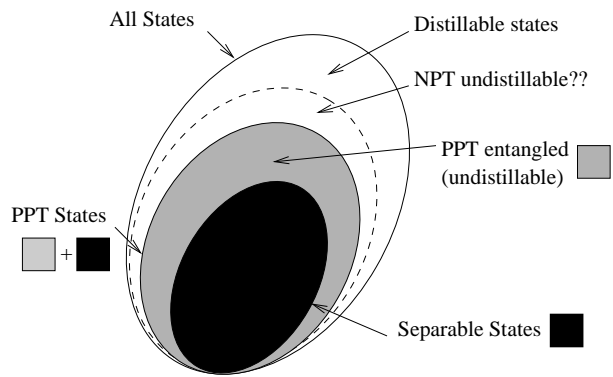,width=8cm}
\end{center}
\caption[]{\small Schematic representation of the set of all states, decomposed into
the various subsets explained in the text}
\label{classif}
\end{figure}

Apart from several examples for bound entangled states with
positive partial transpose we have some evidence that also bound
entanglement, i.e., entanglement that cannot be distilled, of
states with non-positive partial transpose exists: in \cite{nppt}
we study the distillability of a certain class of bipartite
density operators which can be obtained via
     depolarization starting from an arbitrary one.
     This class is a one-parameter family of states
     that consist of a weighted sum
     of projectors onto the symmetric
     and the antisymmetric subspace.
     Our results
     suggest that non-positivity of the partial transpose
     of a density operator is not a sufficient condition for
     distillability, when the dimension of both subsystems is
     higher than two. This conjecture has been found
     independently in \cite{ibm}, and is still an open
     problem. The present understanding of the decomposition of 
     mixed states into separable, undistillable entangled and distillable
     entangled states is shown in figure \ref{classif}.

We have also addressed the distillability and bound
entanglement question in the contexts of infinite
dimensional Hilbert spaces\cite{contvar}. We have introduced
and analyzed the definition of generic bound entanglement for the case
of continuous variables. We have provided
 some examples of bound entangled states for that case, and discussed
 their physical sense in the context of quantum
       optics. We have raised
 the question of whether the entanglement of these states is generic.
As a byproduct, we have obtained a new many-parameter family of
bound entangled states with positive partial transpose in Hilbert
spaces of arbitrary finite dimension.  We have also pointed out
that the ``entanglement witnesses'' (see section \ref{wit}) and
positive maps revealing the corresponding bound entanglement can
be easily constructed.

Furthermore, we have studied how rare separable and non-distillable
states of continuous variables\cite{rare} are. In finite
dimensional Hilbert spaces, we have earlier demonstrated that the
volumes of the set of separable states and PPT entangled states
are both non-zero, and that there exists a  vicinity of the
identity operator that contains separable states only
\cite{volume}. This turned out not to be the case for continuous
variable systems. Also we have proven that the set of
non--distillable continuous variable states is nowhere dense
%
%
in the set of all states, i.e., the states of
infinite--dimensional bipartite systems are generically
distillable. This automatically implies that the sets of separable
states, entangled states with positive partial transpose, and
bound entangled states are also nowhere dense in the set of all
states. All these properties significantly distinguish quantum
continuous variable systems from the spin like ones. The aspects
of the definition of bound entanglement for continuous variables
has also been  analyzed in the context of the theory of Schmidt numbers.
In particular, the main result was generalized to the set of
states of arbitrary Schmidt number  and to the single copy
regime.

\section{Witness operators for the detection of entanglement}
\label{wit}

\subsection{Definition and geometrical interpretation of witness
operators}

A very useful tool to detect entanglement is the so-called 
entanglement witness.  An entanglement witness is an observable
$(\calw)$ which reveals the entanglement (if any) of a given state
$\rho$. This concept, which was introduced and studied in
\cite{horo,witness:Terhal}, reformulates the problem of separability 
in terms of witness operators:

\noindent{\bf {Theorem 3}}
{\em A density matrix $\varrho$ is entangled iff 
there exists a Hermitian operator $\calw$
    with
  Tr$(\calw \varrho) < 0$ and 
    Tr$(\calw \sigma) \geq  0$
  for any separable state $\sigma$.}

We say that  the witness $\calw$ ``detects'' the entanglement of
$\varrho$. The existence of entanglement witnesses is just a
consequence of the Hahn-Banach theorem, that states:
{\em Let $S$ be a convex, compact set, and let
 $\varrho \not\in S$. Then  there exists a hyper-plane that
 separates $\varrho$ from $S$.}

\begin{figure}[ht]
\setlength{\unitlength}{1pt}
\begin{center}
\psfig{file=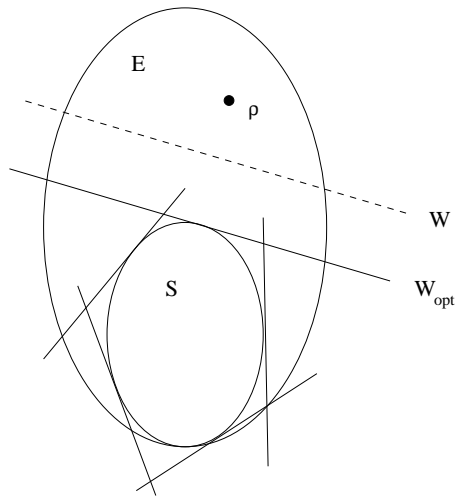,width=6cm}
\end{center}
\caption[]{\small Geometrical picture of entanglement witnesses
and their optimization.}
\label{fig1}
\end{figure}

Figure \ref{fig1} illustrates the concept of an entanglement witness 
$\calw$, represented by a hyper--plane (dashed line)  
that separates the state $\rho$
from the convex compact set $S$. We have also 
depicted in the figure, a optimal entanglement witness ${\cal
W}_{opt}$ (represented by straight line) 
together with other optimal witnesses. Optimal witnesses are 
tangent to the set of separable states (The concept of optimization
will be explained in the next subsection). One can immediately grasp
from the figure, 
that, in order to completely characterize the set
of separable states $S$ one should find all the witnesses tangent 
to $S$. Unfortunately, infinitely many witnesses are needed for such
a task!

Witness operators are also related to maps. Indeed, there is an
isomorphism that connects maps with operators known as
Jamio\l kowski
isomorphism:
each entanglement witness ${\cal W}$ on an $M\times N$
space defines a positive map ${\cal E}$
that transforms positive operators on an $M$ or $N$--dimensional
Hilbert space
into positive operators on
an $M$ or $N$--dimensional space\cite{jamio}.
The maps corresponding to  entanglement witnesses are positive, but
not completely positive (i.e. there is an extension
$\eins\otimes {\cal E}$ which is not positive), and thus
allow to ``detect" the
entanglement of $\rho$.
Entanglement witnesses for PPT states and the corresponding maps
have the property of being {\em non-decomposable}. A witness is
called decomposable iff it can be written in the form
$\calw=P+Q^{T_A}$ with both $P$ and $Q$ positive. Otherwise it is
non-decomposable. Correspondingly a map is decomposable iff it can
be represented as a combination of positive maps and partial
transposition and it is non-decomposable otherwise.
%

\subsection{Results on witness operators}
\label{reswit}

How does one construct an entanglement witness? In \cite{witness}
we provide a canonical form of mixed states in bipartite quantum
systems in terms of a convex combination of a separable state and
an edge state, as defined in section \ref{sep}. We construct
entanglement witnesses for all edge states, and present a
canonical form of {\em non-decomposable} entanglement witnesses
and the corresponding positive maps.
We present a characterization of separable states using a special
class of entanglement witnesses.

An entanglement witness ${\cal W}$ is called {\em optimal}, if
there exists no entanglement witness that detects states further
to the ones detected by ${\cal W}$. Geometrically, this
corresponds to the hyperplane defined by the witness being tangent
to the set of separable states, see figure \ref{fig1}. In
\cite{optwit} we give necessary and sufficient conditions for
entanglement witnesses to be optimal.
 We
     show how to optimize a general witness, and then we
     particularize our results to witnesses
     that can detect PPT entangled states, i.e.
     non-decomposable witnesses.
      This method also permits the systematic construction
     of non-decomposable positive maps.

The tool of witness operators can be applied to give a finer
classification of entangled states by detecting their so-called
{\em Schmidt number}. The Schmidt number of a mixed state was
introduced in \cite{barpaw} as a generalization of the Schmidt
rank for pure states: it characterizes the  maximal Schmidt rank
of the pure states in the ``most simple'' decomposition of $\rho$,
i.e. the one that needs the lowest maximal Schmidt rank. The
definition of the Schmidt number $k$ is given by \be \varrho =
\sum_i p_i\ket{\Psi_i^{r_i}}\bra{\Psi_i^{r_i}} \  , \ \ \
k=\min_{\{ dec\} }( r_{\rm{max}})\ , \ee where $r_i$ denotes the
Schmidt rank of the state $\ket{\Psi_i}$, the minimization is done over
all possible decompositions of $\rho$, and $r_{\rm{max}}=\max_i(r_i)$ is
the maximal Schmidt rank of a given decomposition. In
\cite{schmidt} we investigate the Schmidt number of an arbitrary
     mixed state by constructing a Schmidt
     number witness that detects it. We present a canonical form of
     such witnesses and provide constructive
     methods for their optimization.
     In this context we also find strong evidence
     that all bound entangled states with
     positive partial transpose in two qutrit systems have Schmidt number two.

In the articles summarized above, we were considering bipartite
systems only. Can one use the method of entanglement witnesses for
systems of more than two particles? This question was addressed in
\cite{abls}, where we introduce a classification of mixed
three-qubit states. The case of pure three-qubit states was
studied in \cite{duer}: here the authors show that there exist two
inequivalent classes of states with genuine tripartite
entanglement, the so-called GHZ- and W-states. In \cite{abls} we
define the classes of mixed separable, biseparable, W- and
GHZ-states, which  are successively embedded into each other. We
show that contrary to pure W-type states, the mixed W-class is not
of measure zero. We construct witness operators that detect the
class of a mixed state, and discuss the conjecture that all PPT
entangled states belong to the W-class.
   The classification of three-qubit states into the sets mentioned
   above is schematically shown in figure \ref{fig2}.


Although the concept of entanglement witnesses is born from a mathematical
background, it is by no means purely academic. We have been studying
the possibility of implementing a witness with
only few {\em local} projection
measurements in \cite{detection}.
If  some prior knowledge of the density
     matrix is given,
     which is usually the case in a realistic experiment, one can
      construct a suitable entanglement witness and  find
     its minimal
     decomposition into a {\em pseudo-mixture}
     (i.e. a mixture that contains at least one negative coefficient)
     of
     local projectors, i.e.
\begin{equation}
W=\sum_{i}c_{i}\proj{a_{i}}\otimes\proj{b_{i}}\ ,
\label{deco}
\end{equation}
where the coefficients $c_{i}$ are real and fulfill $\sum_ic_{i}=1$.
     As local projection measurements can be performed
     with present day technology, some simple measurements then
     tell the experimentalist whether his given state
     is indeed entangled.
     The general solution to the optimization problem
     of finding the {\em minimal} number of measurements
     is yet unknown.
     We discuss a realistic example
     for two qubits, and suggest the first method for the detection of
     bound entanglement
     with local measurements.

The tool of witness operators is not only useful for
addressing the separability problem, but also  for
studying the distillability problem:
In \cite{activable}
we introduce a formalism that connects entanglement
witnesses and the distillation and activation
     properties of a state. We apply this formalism to two
     cases: First, we rederive the results presented in
     \cite{egge}, namely that one
     copy of any bipartite state with non--positive partial
     transpose  is either distillable, or activable.
      Second, we show that there exist three--partite NPPT
     states, with the property that two copies can neither
     be distilled, nor activated.

 Finally, an overview of our programme that investigates
quantum correlations and entanglement in terms of
     convex sets is given in
    \cite{reflections}. There we present a unified description of optimal
     decompositions of quantum states and the optimization
     of witness operators that detect whether a given state
     belongs to a given convex set. We illustrate this
     abstract formulation with several examples, and discuss
     relations between optimal entanglement witnesses
     and n-copy non-distillable states with non-positive
     partial transpose.

\section{Quantum correlations in systems of fermionic and bosonic states}
\label{fer}

The notion of entanglement discussed in the previous sections applies to situations
where the parties are separated by macroscopic distances. Various mechanisms to create
entanglement or to perform quantum gate operations, e.g.~in the context of
quantum dots \cite{SLM:00} or neutral atoms in optical microtraps \cite{traps},
however require a direct interaction at short distances between
indistinguishable particles.
We have developed a framework to study quantum correlations in such situations where
the bosonic or fermionic character of indistinguishable particles become important.
We have furthermore described a possible implementation of a quantum logic gate
for neutral atoms in optical microtraps and studied bosonic correlations
in this case.
%

\subsection{What is different with indistinguishable particles?}

To illustrate the consequences of indistinguishability consider two fermions
located
in a double well potential as a schematic model of electrons in quantum dots
and assume
the qubit to be implemented in the spin degree of freedom. Let the initial
situation
be such that each well contains one electron.
Even if they are prepared completely independently, their pure quantum state
has to be written in terms of Slater determinants in order to respect the
indistinguishability. Operator matrix elements between such Slater
determinants contain terms due to the antisymmetrization, but
if the spatial wavefunctions of electrons located in different wells have only
vanishingly small overlap, then the matrix elements will tend
to zero for any physically meaningful operator. This situation is generically
realized if the supports of the single-particle wavefunctions are essentially
centered around locations being sufficiently apart from each other, or the
particles are separated by a sufficiently large energy barrier.
In this case the antisymmetrization has no physical effect and for all
practical
purposes it can be neglected.

If the two wells are moved closer together, or the energy barrier is lowered,
such
that the electrons are no longer completely localized in one well,
then the fermionic statistics is clearly essential and the two-electron
wave-function has to be antisymmetrized.
Note that in this situation the space of states written in terms of
single-particle states
no longer has a tensor product structure because the actual state space is
just a subspace
of the complete tensor product. As a consequence of this fact
any antisymmetrized
state formally resembles an entangled state although these correlations are
not useful
as individual particles cannot be accessed. To emphasize this fundamental
difference
between distinguishable and indistinguishable particles, we will
use the term \emph{quantum correlations} to characterize \emph{useful}
correlations in systems of indistinguishable particles as opposed to
correlations
arising purely from their statistics.

We remark that there are different possible ways to quantify
quantum correlations. An approach which can be seen as
complementary to the one which we will describe here was discussed
by Zanardi \cite{Zan:01} who ignored  the original tensor product
structure through partitioning of the physical space into
subsystems and introducing a tensor product structure in terms of
modes. The entangled entities then are
 no longer particles but modes.

\subsection{Results on quantum correlations for indistinguishable particles}

Let us consider the case of two identical fermions sharing an
$N$-dimensional single-particle space ${\cal H}_{N}$. The total
Hilbert space is ${\cal A}({\cal H}_{N}\otimes{\cal H}_{N})$ where
${\cal A}$ denotes the antisymmetrization operator. A general state vector
can be written as
\be
|w\rangle=\sum_{i,j=1}^{N}w_{ij}f^{\dagger}_{i}f^{\dagger}_{j}|\0\rangle
\ee
with fermionic creation operators $f^{\dagger}_{i}$ acting on the vacuum
$|\0\rangle$. The antisymmetric coefficient
matrix $w_{ij}$ fulfills the normalization condition
${\rm tr}\left( w^* w\right)=-1/2\,$.
Under a unitary transformation of the single-particle space,
$f^{\dagger}_{i}\mapsto\sum_j U_{ji}f^{\dagger}_{j}\,$, $w$ transforms as
$w\mapsto UwU^{T}\,$.

\noindent{\bf {Theorem 4}}
For every pure two-fermion state $\ket{w}$
there exists a unitary transformation of the single particle space such that
in the new basis of creation operators $\fc{i}$ the state is of the form
\be\label{eqn:slaterDecomp}
\ket{w}=2\sum_{k=1}^{m}z_k\fc{2k}\fc{2k-1}\ket{\0}
\ee
with $2\cdot m\leq N$ and $z_k$ real and positive.

Each term in this decomposition corresponds to an elementary
Slater determinant which is an analogue of a product state in
systems consisting of distinguishable parties. Thus, when
expressed in such a basis, $|w\rangle$ is a sum of elementary
Slater determinants where each single-particle basis state enters
at most one term. In this basis the number $m$ of Slater
determinants is furthermore minimal and these Slater determinants
are thus the analogues of the products states occurring in the
Schmidt decomposition of a bi-partite state of distinguishable
particles. Therefore we call $m$ the
{\em fermionic Slater rank} of
$|w\rangle$ \cite{2fermion}, and an expansion of the form
(\ref{eqn:slaterDecomp}) a
{\em Slater decomposition} of $|w\rangle$. For bosons there exists
a similar expansion in terms of elementary two-boson Slater
permanents representing doubly occupied states \cite{identical}.

For two fermions the smallest single-particle space allowing for non-trivial
correlations is four-dimensional. In this case a quantity analogous to
the
concurrence introduced by Wootters as an entanglement measure for
two distinguishable qubits \cite{wooters} can be constructed in
the following way:

\noindent{\bf {Theorem 5}} Let $\ket{w}$ be a two-fermion state in
a four-dimensional single-particle space. Then the concurrence
$\calc(\ket{w})$, defined as \be
\calc(\ket{w})=\Big|\frac{1}{2}\sum_{i,j,k,l=1}^{4}\epsilon^{ijkl}w_{ij}w_{kl}\Big|,
\ee ($\epsilon$ is the fully antisymmetric unit tensor) has the
following properties: (i) $\calc(\ket{w})$ is invariant under
unitary transformations of the single-particle space, (ii)
$0\le\calc(\ket{w})\le1$ and (iii) $\calc(\ket{w})=0$ iff
$\ket{w}$ has Slater rank one and $\calc(\ket{w})=1$ iff $\ket{w}$
has maximal Slater rank, i.e. Slater rank two.

The concurrence $\calc(\ket{w})$ thus fully characterizes quantum correlations
in the case of pure states and for $N=4$. For mixed two-fermion states in a
four-dimensional single-particle space
characterized by a density matrix $\rho$ a
{\em Slater number} can be defined similar to the Schmidt number
for mixed states of two qubits as the maximal Slater rank of a
decomposition of $\rho$ into pure states minimized over all
decompositions. Also we can define the mixed state concurrence as
\be \calc(\rho)=\inf_{\{p_i,\ket{w_i}\}}\left\{\sum_i
p_i\calc(\ket{w_i})\right\} \ee where the infimum is taken over
all decompositions of $\rho$. With this definition we find:

\noindent{\bf {Theorem 6}}
Let $\rho=\sum_i p_i\proj{w_i}$ be a mixed two-fermion state. Define
$\ket{\widetilde w_i}=\sum_{i,j,k,l=1}^4\epsilon^{ijkl}w_{kl}\fc{i}\fc{jl}\ket{\0}$
and $\tilde\rho=\sum_i p_i\proj{\widetilde w_i}$
and let $\lambda_i$ be the real and non-negative eigenvalues of $\rho\tilde\rho$
in descending order of magnitude. Then
\be
\calc(\rho)=\max(0,\lambda_1-\sum_{i=2}^6\lambda_i)
\ee
and $\rho$ has Slater number one iff $\calc(\rho)=0$, i.e iff
$\lambda_1\leq\sum_{i=2}^6\lambda_i$.

Notice that in a similar way the concurrence can be defined and
calculated for pure and mixed states of two bosons in a
two-dimensional single-particle space.

For higher-dimensional single-particle spaces there exist
necessary and sufficient criteria to determine the Slater rank of
pure fermionic and bosonic states by contracting their coefficient
matrix $w$ with the $\epsilon$-tensor \cite{identical}. These
become only necessary criteria when applied to mixed states and
apparently a full and explicit characterization of
higher-dimensional two-boson and two-fermion mixed states is not
possible. Furthermore for the case of more than two particles a
straight-forward generalization of the Slater decomposition cannot
be given. This is again similar to the case of more than two
qubits where a Schmidt decomposition of a general pure state does
not exist \cite{threequbit}. Consider for example states of three
fermions in a six-dimensional Hilbert space. It is in general not
possible to find a unitary transformation of the single-particle
space that brings a given state to a form $\ket{w}\propto
z_1\fc{1}\fc{2}\fc{3}\ket{\0}+z_2\fc{4}\fc{5}\fc{6}\ket{\0}$,
which would be the analogue of the two-fermion Slater
decomposition. There however exist criteria to identify pure
uncorrelated states, i.e.~states that can be written as a single
Slater determinant \cite{identical}.

Finally we notice that for the case of two fermions or bosons in
higher-dimensional
single particle spaces ($N>4$ for fermions, $N>2$ for bosons) the concepts
of witnesses can be applied \cite{2fermion,identical}. As
explained in section \ref{wit} for the case of distinguishable particles,
$k$-edge states can be introduced as states that become non-positive when
$\epsilon\proj{w^{<k}}$ is subtracted for some state $\ket{w^{<k}}$ of
Slater rank $<k$. Then
fermionic and bosonic $k$-Slater witnesses can be defined that
detect states of Slater number $k$. These witnesses can
furthermore be optimized as demonstrated in section \ref{wit}.

\subsection{Implementation of an entangling gate with bosons}



In \cite{traps} we investigate quantum computation with bosonic neutral 
atoms in optical microtraps \cite{birkl}. In contrast to other methods with
the qubit being implemented in an internal degree of freedom, we study the case where the
qubit is implemented in the motional state of the atoms, i.e., in the
two lowest vibrational states of each trap. The quantum gate operation
is performed by adiabatically approaching two traps each containing one particle
such that tunneling and cold collisions occur and thus the bosonic character of the atoms 
is important. We especially address the implementation of a $\sqrt{\text{SWAP}}$-gate, i.e.,
a two-qubit gate that transforms states $\ket{0}_A\ket{1}_B$ and $\ket{1}_A\ket{0}_B$
to maximally entangled states while leaving $\ket{0}_A\ket{0}_B$ and $\ket{1}_A\ket{1}_B$
unchanged. The fidelity of the gate operation is evaluated as a function of the
degree of adiabaticity in moving the traps and for rubidium atoms in state-of-the-art optical
microtraps we obtain gate durations in the range of a few tens of milliseconds.
Taking into account error mechanisms like spontaneous scattering of photons we calculate
error rates of the gate operation and show that proof-of-principle
experiments should be possible.  

\section{Summary}
The characterization and classification of entangled states
is a very challenging open problem of modern quantum theory.
We have presented some approaches and partial solutions
to this problem. The methods we used are  the optimal decomposition
of a given state into a separable  
and
an entangled state, and the tool of witness operators. We have
summarized  various advances in the separability and distillability
problem, and addressed the question of experimental implementation
of witness operators. However, many open questions still remain to be
solved.

 \section{Acknowledgments}
 We wish to thank  A. Ac\'\i n,
 G. Birkl, I. Cirac, W. D\"{u}r, A. Ekert, G. Giedke,
 P. Horodecki, S. Karnas,
 B. Kraus, M. Ku\'s, D. Loss,
 C. Macchiavello, A. Peres, J. Schliemann,  G. Vidal  and X. Yi
 for collaboration and discussions.
 This work has been supported by the DFG-Schwerpunkt
 ``Quanteninformationsverarbeitung''.

\end{document}